\documentclass[conference]{IEEEtran}
\IEEEoverridecommandlockouts
\usepackage{cite}
\usepackage{amsmath,amssymb,amsfonts}
\usepackage{algorithmic}
\usepackage{graphicx}
\usepackage{textcomp}
\usepackage{xcolor}
\def\BibTeX{{\rm B\kern-.05em{\sc i\kern-.025em b}\kern-.08em
    T\kern-.1667em\lower.7ex\hbox{E}\kern-.125emX}}
\begin{document}

\title{Towards a Universal Features Set for IoT Botnet Attacks Detection

\thanks{[© 20xx IEEE. Personal use of this material is permitted. Permission from IEEE must be obtained for all other uses, in any current or future media, including reprinting/republishing this material for advertising or promotional purposes, creating new collective works, for resale or redistribution to servers or lists, or reuse of any copyrighted component of this work in other works.]}

\author{\IEEEauthorblockN{Faisal Hussain, Syed Ghazanfar Abbas}
\IEEEauthorblockA{\textit{Al-Khawarizmi Institute of Computer} \\
Science (KICS) Lahore, Pakistan \\
faisal.hussain.engr@gmail.com, \\
ghazanfar.abbas@kics.edu.pk}
\and
\IEEEauthorblockN{Ubaid U. Fayyaz, Ghalib A. Shah}
\IEEEauthorblockA{\textit{Al-Khawarizmi Institute of Computer} \\
Science (KICS) Lahore, Pakistan \\
ubaid@uet.edu.pk, \\
ghalib@kics.edu.pk}
\and
\IEEEauthorblockN{Abdullah Toqeer, Ahmad Ali}
\IEEEauthorblockA{\textit{Al-Khawarizmi Institute of Computer} \\
Science (KICS) Lahore, Pakistan \\
abdullah.toqeer@kics.edu.pk, \\
ahmad.ali@kics.edu.pk}
}
}

\maketitle
\begin{abstract}
The security pitfalls of IoT devices make it easy for the attackers to exploit the IoT devices and make them a part of a botnet. Once hundreds of thousands of IoT devices are compromised and become the part of a botnet, the attackers use this botnet to launch the large and complex distributed denial of service (DDoS) attacks which take down the target websites or services and make them unable to respond the legitimate users. So far, many botnet detection techniques have been proposed but their performance is limited to a specific dataset on which they are trained. This is because the features used to train a machine learning model on one botnet dataset, do not perform well on other datasets due to the diversity of attack patterns. Therefore, in this paper, we propose a universal features set to better detect the botnet attacks regardless of the underlying dataset. The proposed features set manifest preeminent results for detecting the botnet attacks when tested the trained machine learning models over three different botnet attack datasets. 

\end{abstract}


\section{Introduction}
We are living in the age of the internet where the real-world things are now becoming smart, intelligent, connected to the internet, and capable of interacting with one another without human intervention in order to ease and lavish the human life \cite{xia2020modeling}. The connection of real-world objects with the internet inaugurated the concept of internet of things (IoT). IoT is a communication paradigm where a wide range of daily life objects is connected via internet. With the evolution of IoT, smart devices stepped into our daily life. IoT has revolutionized many innovative applications of smart technology like smart home, smart office, smart grid, smart healthcare, smart agriculture, smart transportation, smart city, etc. These applications would have a great influence in our lives thus human life will become dependent on smart devices \cite{ghazanfar2020iot}.

Despite the valuable revolution, security is the major concern of IoT \cite{hossain2019application}. The IoT vendors are focusing on increasing the device features instead of making it secure to bring their product into the market to get more profit as early as possible \cite{alladi2020consumer}. Hence, the security of IoT devices is ignored. The security breach of IoT devices would have a severe effect on our lives due to much utilization of smart devices in our daily life necessities. IoT devices and applications are also being used in critical infrastructures like power plants, water plants, nuclear plants, transportation systems, healthcare systems, etc., to make them remotely accessible and more efficient. The burgeoning applications of IoT devices in critical infrastructures are also increasing the potential risk of cyber-security attacks since the IoT devices inherent some serious cyber-security problems with them \cite{OWASP}. Hence, any cyber-attack at these critical infrastructures may cause catastrophic results if the IoT devices and applications are not secured properly.

The IoT devices inherit some serious cyber-security problems \cite{OWASP} like weak security configurations, weak or hardcoded passwords, etc. This is due to the subpar focus of IoT vendors on strong security measures \cite{ahmed2019protecting} \cite{pour2020data}. These security pitfalls of IoT devices have made it easy for hackers to take over IoT devices and use them for malicious activities like botnet attacks \cite{bertino2017botnets}. Botnets are the connected network of malware-infected devices which are remotely controlled by command \& control servers \cite{xia2020modeling}. The attackers use the botnet for malicious activities like sending spam emails, click fraud, launching distributed denial of service (DDoS) attacks to chop down a web-service, etc. Botnets existed for many years, but with the proliferation of insecure IoT devices, botnets have become larger, complex and dangerous.

At present, Botnet attacks have become a serious threat to the whole internet \cite{xia2020modeling}. The escalating adoption of insecure IoT devices have made it an easy job for the attackers to exploit the IoT devices, and make them a part of the botnet army to conduct the large scale malicious activities \cite{xia2020modeling}. The botnet attacks are not only catastrophic for IoT device users but also for the rest of the world since these botnet caused ever large and devastating DDoS attacks at the famous web service providers like GitHub \cite{gitAttack}, Krebs on Security, etc., in recent years \cite{xia2020modeling}. Mirai is the pioneer example of ever large and powerful DDoS attack till 2016 that occurred through a botnet of more than 2000,000 IoT devices \cite{pour2020data}. In Mirai, the hackers exploited the open ports, default or hard-coded credentials of thousands of IoT devices including surveillance cameras, baby toys, wireless printers, etc., and made them a part of botnet \cite{ahmed2019protecting}. Likewise, Echobot attack recently exploited more than 20 unique IoT vulnerabilities and compromised millions of IoT devices \cite{pour2020data}. No matter, the present firewall and IDS technologies are quite mature but it is inadequate for IoT Systems due to their versatile traffic patterns, communication protocols, etc. \cite{pour2020data} \cite{zarpelao2017survey}. Therefore, it is the need of the hour to design intelligent IoT-specific security solutions and integrate them with the existing security infrastructure in order to better protect the IoT devices from botnet attacks.

So far, many studies are suggesting different machine learning (ML) based solutions to detect the botnet, DDoS attacks but their performance is limited to a specific dataset on which they are trained. This is because the features used to train an ML model work good on one certain botnet dataset, but does not perform well when tested on other datasets due to the diversity of attack patterns. Since, the performance of ML models greatly rely on features set used for training \cite{barnes2017assessing}. Therefore, in this paper, we propose a universal features set that better helps the ML models in discriminating the botnet attacks from the normal traffic regardless of the underlying dataset. Thus, the key objective of this research is to extrapolate a universal features set which performs well for detecting botnet attacks irrespective of a certain dataset. The key contributions of this work are as follows:

\begin{itemize}
    \item We proposed a universal features set that better helps the machine learning algorithms in discriminating the botnet attacks from the normal traffic irrespective of the underlying dataset.  
    \item We compared the performance of four commonly used machine learning classifiers across the proposed features set over three different botnet attack datasets.   
    \item	Furthermore, we analyzed the performance of the proposed features set over three publicly available datasets in order to validate the applicability and efficiency of the proposed universal features set. 
\end{itemize}

The rest of the paper is organized as follows: Section II presents a review of some existing work for botnet attack detection. Section III describes the entire methodology from the data acquisition to botnet attack detection. Furthermore, the methodology that is followed to propose a universal features set for detecting the bot attacks, is also explained in this section. Section IV discusses the features selection results and the botnet attacks detection results across three different botnet attack datasets. Lastly, Section V concludes the paper.

\section{Related Work} 

To date, many approaches have been proposed for detecting the botnet attacks. These approaches can be broadly categorized into two types, i.e., the flow-based approach and graph-based approach. In a flow-based method, botnets are detected by extracting the network flow-based features. While in the graph-based method, botnets are detected based upon the graphical communication structure among different nodes.

Frank \textit{et al.} \cite{frank2018protecting} proposed a methodology to prevent IoT devices from being part of Mirai botnet. The proposed approach consists of two scripts, i.e., hardening script and detection script which are installed on end device in order to protect it from Mirai botnet. The hardening script is for preventing the IoT devices from bot attacks whereas the detection script is proposed to recognize whether the underlying IoT device is part a Botnet or not.

Ryu \textit{et al.} \cite{ryu2018comparative} analyzed the effect of ensembling machine learning algorithms with a neural network for botnet detection. They ensembled decision tree and Naive Bayes classifiers with a neural network and concluded that the proposed ensembling method can better detect botnet attacks in network traffic as compared to individual classifiers. 
Keisuke \textit{et al.} \cite{kato2017development} developed a network anomaly detection system using a Gaussian model to detect the bot attacks. They applied principle component analysis (PCA) to downsize the dataset features and classified the network traffic as normal or attack by using a Gaussian mixture model.
\begin{figure*}[t]
 \centering
  \includegraphics[width=\linewidth]{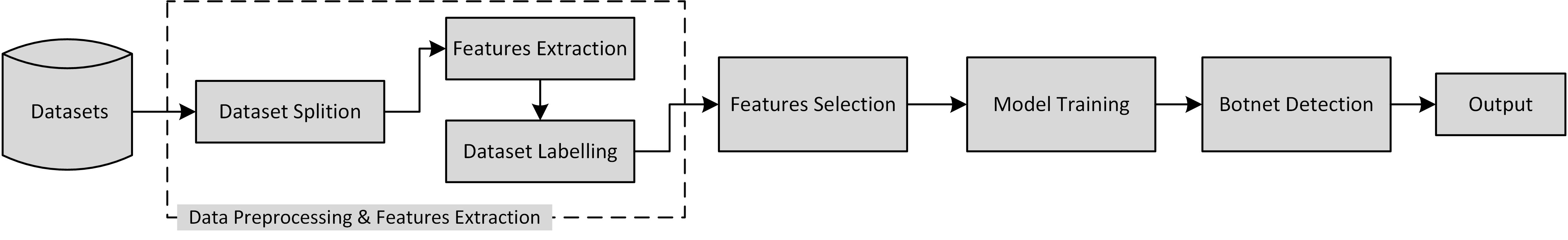}
  \caption{Methodology for Botnet Attacks Detection}
  \label{fig:methodology1}
\end{figure*}

Sofiane \textit{et al.} \cite{lagraa2017botgm} proposed a graphical method to detect the botnet attacks. They first build a graph of flow sequences based upon the source and destination IP addresses of hosts and then applied the unsupervised model to detect the outliers in the traffic to detect the botnet behaviour. Similarly, Wang \textit{et al.} \cite{wang2017botnet} suggested a two-phase model for botnet detection. In the first phase, they used a flow-based approach to detect a botnet attack.  In the second phase, they used a graph-based technique to trace the associated nodes of botnets communication. Likewise, Abbas Abou \textit{et al.} \cite{daya2019graph} designed a versatile graph to detect the botnet attacks. The authors build a communication graph by representing hosts as nodes and communications between them as vertices. Afterwards, they extracted the graphical features and applied different machine learning techniques to better detect the botnets.

Michal \textit{et al.} \cite{piskozub2019malalert} investigated the impact of different types of malwares like adware, trojans, botnets, etc., and proposed a system to detect the malwares based upon network flow-level features. The authors extracted 441 statistical features from network flow and then used mutual information to select the most significant features. Finally, they used Random Forest classifier and predicted the malware types with 90\% F1-score. Tao \textit{et al.} \cite{chen2020novel} identified that C\&C channel is a compulsory part of botnet attack. Therefore, it is crucial to detect and blocking the C\&C channel to stop the botnet attacks. They proposed an ensemble anomaly detection technique which composed of a one-class SVM and convolutional autoencoder. The trained the ensemble model only at normal traffic in order to reduce the false alarm rate and to better detect the botnet. Similarly, Blaise \textit{et al.} \cite{blaise2020botfp} analyzed the botnet behaviour using unsupervised techniques. The authors calculated the frequency distribution of protocol features to characterize the host behaviour and used clustering techniques to identify the bots from the normal hosts.


Most of the above studies proposed different machine learning techniques to detect botnet attacks. There exist some works like \cite{letteri2019feature, guerra2019hybrid}, etc., that proposed hybrid feature selection methods to better detect IoT botnet attacks. However, the performance of these techniques is limited to a specific dataset on which they are trained. This is because the features used to train a machine learning model on one botnet dataset, does not perform well on other datasets due to the diversity of attack patterns. Therefore, in this paper, we proposed a universal features set which perform well for detecting botnets on all datasets.

\section{Methodology}
To better detect the botnet attacks irrespective of a dataset, this paper proposes a universal features set which is extrapolated based upon Logistic Regression algorithm and frequency counting technique. The entire methodology from the data acquisition to botnet attack detection consists of six major steps as shown in Fig. \ref{fig:methodology1}. These steps include data acquisition, data pre-processing, features extraction, features selection, model training, and botnet attacks detection. These steps are explained in the following subsections:

\subsection{Data Acquisition}

The data acquisition is the premier step for training a machine learning model. Traditionally, two methods are followed to acquire the data. One way is to get the data from an existing real-time environment or to set up a real-time environment in a laboratory in order to get the data for training a machine learning model. Acquiring data from a real-time environment is not possible for everyone since there exist some privacy issues due to which organizations do not share their data. Similarly, establishing a real-time environment in a laboratory is also a tough job since it consumes time, resources, and money. 

In order to get rid of these concerns, most of the researchers follow the other way, i.e., use some publicly available datasets for data acquisition. In our case, we also followed the second way and acquired data from three publicly available datasets which include CICIDS2017 \cite{sharafaldin2018toward}, CTU-13 \cite{garcia2014empirical} and IoT-23 \cite{iot-23}. The CICIDS2017 \cite{sharafaldin2018toward} dataset contains up-to-date real-time packet captures of different attacks including the DDoS attack. The CTU-13 \cite{garcia2014empirical} dataset contains traditional botnet packet captures while the IoT-23 \cite{iot-23} contains IoT botnet packet captures. All these datasets exist in packet capture (.pcap) file format along with labels given in text files. 

From the above-mentioned datasets, we only considered the .pcap files that contain botnet attack traffic and normal network traffic. Thereby, we considered all the .pcap files from CTU-13 \cite{garcia2014empirical} dataset since it contains only botnet attacks and normal network traffic. In case of, CICIDS2017 \cite{sharafaldin2018toward} dataset, we only considered the .pcap files having DDoS attack and normal traffic .pcap files. Finally, from IoT-23 \cite{iot-23} dataset, we only considered the famous Mirai botnet attack and normal network traffic .pcap files.

\subsection{Data Preprocessing \& Features Extraction}
The acquired datasets were in .pcap format while we need features to train a machine learning algorithm. Therefore, we first need to extract the features from dataset .pcap files to proceed the further steps. For this purpose, we used an open-source tool named as CICFlowmeter \cite{CICFLOWMETER} to extract the features from .pcap files. The CICFlowmeter tool extracts more than 80 features from a given .pcap file, however, there is a limitation of this tool that it cannot handle a file of size greater than 100MB. Moreover, the acquired datasets contain .pcap files having size more than 100MB.

\textbf{Dataset Splition:} Since the CICFlowmeter \cite{CICFLOWMETER} cannot process the .pcap files having size greater than 100MB, so, we need to do some data pre-processing before extracting the features. In this regard, we used tcpdump utility \cite{tcpdump} which splits the larger .pcap files into the smaller files of a given size. So, by using tcpdump utility \cite{tcpdump}, we converted the .pcap files of the datasets having size larger than 100MB into smaller .pcap files of size less than equal to 100MB. 

\textbf{Features Extraction:} After splitting the larger .pcap files, we then passed the smaller splitted .pcap files to CICFlowmeter \cite{CICFLOWMETER} in order to extract the features. The CICFlowmeter \cite{CICFLOWMETER} extracts more than 80 features (across each network flow) from a given .pcap file which can be categorized into two types, i.e.,  static and dynamic features. The static features include Flow ID, Source IP, Source port, Destination IP, Destination port, and protocol name. While the dynamic features include about 80 flow features like flow Duration, Number of packets, Number of bytes, etc. The description of these features can be found at \cite{cicflowmeter_features}. All these features are extracted across each flow of a given .pcap file and saved inside a .csv file. However, these obtained .csv files are unlabelled.

\textbf{Dataset Labelling:} The datasets labels were given in text files. In order to label these .csv files, we used SQL server. We imported both the .csv files (having size $\leq$ 100MB) and text files and compared the five tuples, i.e, source IP, source port, destination IP, destination port and protocol to identify the normal and malicious flow as defined by dataset providers. Based upon the matching of five-tuple for each flow, the .csv files were labelled with the label that exists across the similar tuple in the text file. In this way, all the datasets were labelled.

\begin{figure}[!b]
 \centering
  \includegraphics{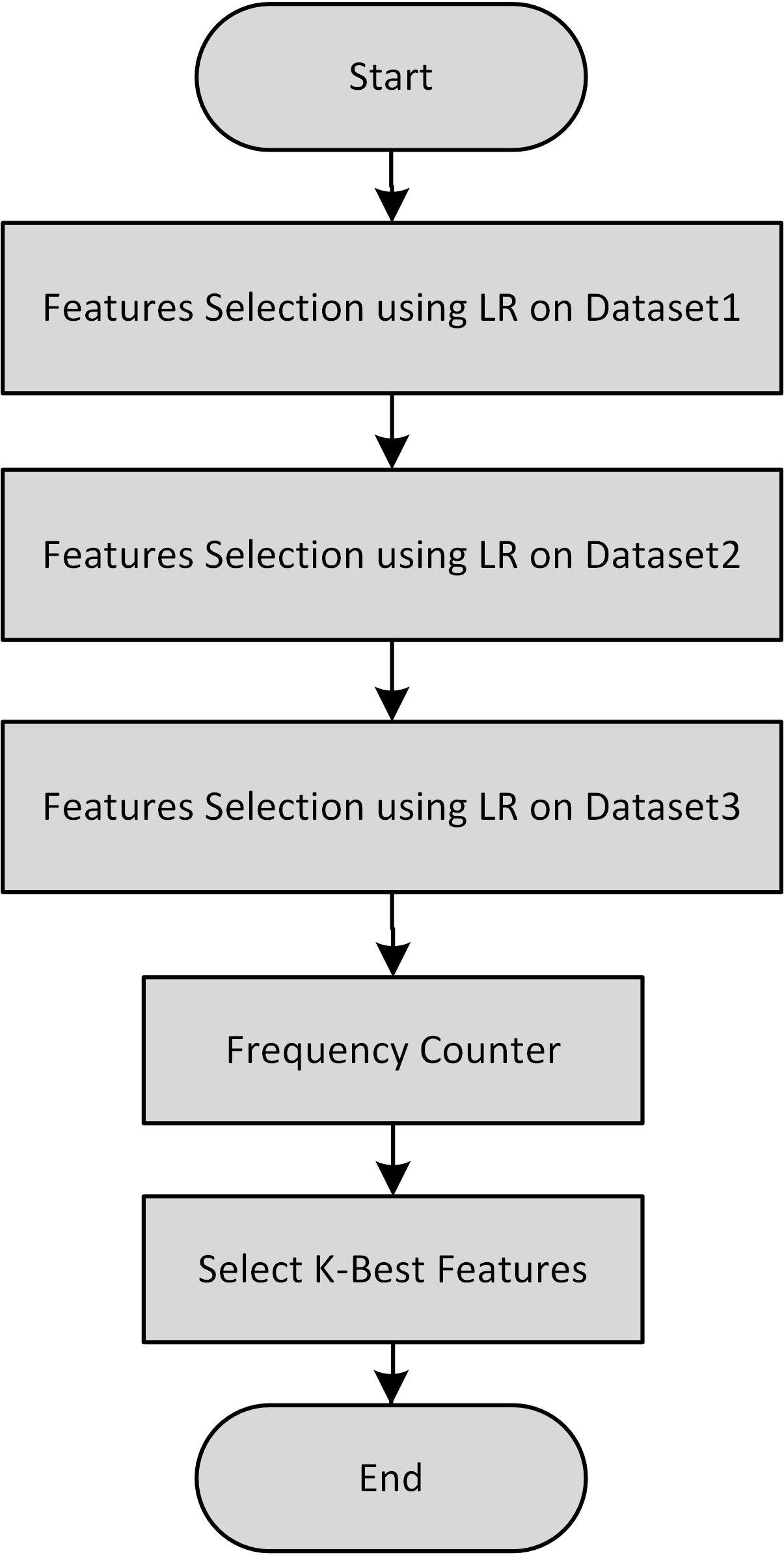}
  \caption{Steps for Extrapolating the Universal Features Set for Botnet Detection}
  \label{fig:methodology2}
\end{figure}

\subsection{Features Selection}
\label{sec:FeaturesSelection}

The selection of features plays a significant role in the performance of a machine learning model \cite{barnes2017assessing}. Usually, the researchers apply feature selection technique on a dataset and train the machine learning models over the selected features. Apparently, in this way, the machine learning models perform good on one dataset but do not perform well when tested on different dataset \cite{stamos2015learning}. Therefore, we propose a universal feature set that can efficiently support the machine algorithms for better detection of the botnet attacks irrespective of a dataset. 

Fig. \ref{fig:methodology2} shows the steps followed to extrapolate a universal features set for botnet attack detection. We used logistic regression (LR) algorithm for feature selection because it is simple, fast and less complex as compared to other techniques \cite{dedeturk2020spam}. 
After pre-processing the Dataset1, we applied the LR algorithm to get the 10 most significant features from Dataset1. Similarly, we applied the LR algorithm on both Dataset2 and Dataset3 to get the 10 most significant features from each dataset. These features are enlisted in Table \ref{featuresList} and discussed in Section \ref{sec:FSResults}.

After getting the significant features list from each dataset, we performed a frequency analysis in order to observe which features are more frequently selected for detecting the Botnet attacks across all three datasets. Based, upon this analysis, we found six most frequently selected features that are selected across all three datasets. Hence, we grouped and named these features as universal features set. Finally, we selected these universal features from each dataset in order to train the machine learning models for detecting the botnet attacks.

\subsection{Model Training}
After the feature selection, the next step is to train the machine learning models for detecting the botnet attacks. For this purpose, we selected the above-mentioned six features individually from all three datasets, i.e., CICIDS2017 \cite{sharafaldin2018toward}, CTU-13 \cite{garcia2014empirical}, and IoT-23 \cite{iot-23}. After that we split each dataset individually into training and testing set with a proportion of 80:20, i.e., 80\% data selected randomly for training and 20\% data selected randomly for testing purpose and passed it for training the four commonly used machine learning algorithms to detect botnet and DDoS attacks. These machine learning algorithms include Naïve Bayes (NB), K-Nearest Neighbours (KNN), Random Forest (RF), and Logistic Regression (LR). We used python scikit-learn library to initialize the training of machine learning models for each dataset individually by feeding the above-mentioned six features extracted from each dataset.

\subsection{Botnet Attack Detection}
Once the model is trained for detecting the botnet attacks, the final step is to test the performance of the trained model over unseen data. 
As mentioned earlier, we selected 20\% data for testing the performance of the trained model. So, in this stage, we tested each trained over 20\% unseen data and calculated its performance across four parameters. These parameters are defined in the following section.
 

\begin{table}[t]
\renewcommand{\arraystretch}{1.3}
\centering
\caption{10 Best Features selected from three Datasets using Lositic Regression Algorithm}
\label{featuresList}
\begin{tabular}{p{2.3cm} p{2.3cm} p{2.9cm}}
\hline
\textbf{IoT-23 \cite{iot-23}}  & \textbf{CTU-13 \cite{garcia2014empirical}} & \textbf{CICIDS-17 \cite{sharafaldin2018toward}}  \\  \hline
Pkt Len Mean & Init Bwd Win Byts & Inbound	 \\ 
Bwd Pkt Len Min & Bwd Pkts/s & Average Packet Size	 \\ 
Pkt Len Min & Flow Pkts/s & Avg Fwd Segment Size	 \\ 
Pkt Size Avg & Fwd Pkts/s & Fwd Packet Length Mean	 \\ 
Bwd Header Len & Pkt Len Mean & Fwd Packet Length Min	 \\ 
Bwd IAT Max & Pkt Size Avg & Min Packet Length	 \\ 
Bwd Pkt Len Mean & Active Mean & Packet Length Mean	 \\ 
Flow Byts/s & Active Min & URG Flag Count	 \\ 
Flow IAT Max & Bwd IAT Min & Down/Up Ratio	 \\ 
Fwd Pkt Len Mean & Down/Up Ratio & Bwd Packet Length Min	\\ \hline

\end{tabular}
\end{table}

\section{Results and Discussions}
\subsection{Features Selection Results}
\label{sec:FSResults}
As mentioned in Section \ref{sec:FeaturesSelection} that we first selected the top 10 most significant features from all three datasets, i.e., CICIDS2017 \cite{sharafaldin2018toward}, CTU-13 \cite{garcia2014empirical}, and IoT-23 \cite{iot-23}. Table \ref{featuresList} shows the list of 10 best features for detecting Botnet attacks obtained by applying LR algorithm across each dataset.

\begin{figure}[t]
 \centering
  \includegraphics[width=8.5cm, height=6.2cm]{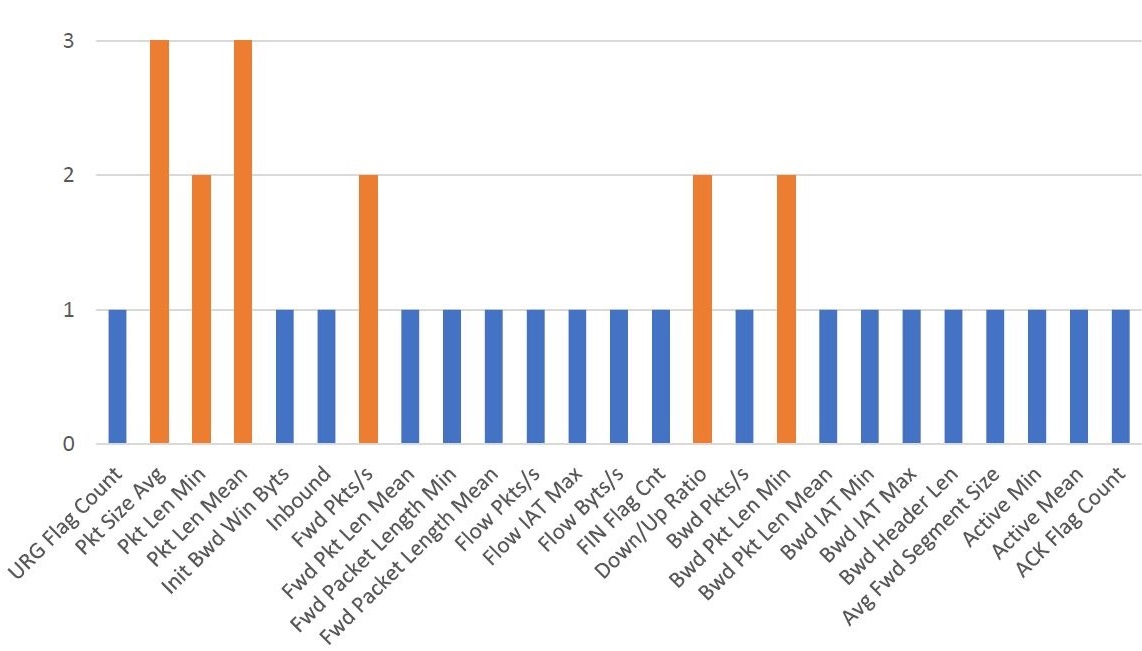}
  \caption{Features Selection Count across Three Datasets}
  \label{fig:Features}
\end{figure}

Afterwards, we applied frequency counting method in order to get a universal features set that can efficiently support the machine algorithms for better detection of the botnet attacks irrespective of a dataset. The frequency counting method calculated the frequency score for each selected feature across all three datasets. Finally, we identified the six most significant features for botnet attacks detection based upon the frequency count of each feature. Fig. \ref{fig:Features} shows the overall features selection count across all three datasets. Based, upon this analysis, we found six most frequently selected features that are selected
across all three datasets. We grouped and named these features as universal features set. These features are highlighted with orange colour bars in Fig. \ref{fig:Features}. We used the proposed universal features set to train and validate the performance of machine learning models for botnet attacks detection across three datasets.


\subsection{Botnet Attacks Detection Results}

As mentioned earlier, after identifying the universal features set for botnet attacks detection, we trained four commonly used machine learning models, i.e., NB, KNN, RF, and LR for detecting different types of botnet attacks across three datasets.

In order to evaluate the performance of the trained models, we calculated four performance matrices. These metrics include accuracy, precision, recall, and F1-measure. 

\textbf{Accuracy} - {} It is defined as the ability of the system to correctly classify the botnet attacks as an “attack” and normal traffic as a “normal flow”. It tells about the ratio of correct predictions with respect to all samples. Mathematically, it is expressed in equation \ref{eq:accuracy}:

\begin{equation}
\label{eq:accuracy}
\textit{Accuracy} = \frac{TP+TN}{TP+FN+TN+FP} \times 100
\end{equation}

\textbf{Precision} - {} It tells about how many of the predicted botnet attacks were correct. It is a ratio between the correctly predicted attacks (i.e., TP) and the actual results (i.e., TP+FP). Mathematically, it is expressed in equation \ref{eq:precision}: 

\begin{equation}
\label{eq:precision}
\textit{Precision} = \frac{TP}{TP+FP} \times 100
\end{equation}

\textbf{Recall} - {} It is defined as the ability of the system to correctly detect the botnet attack upon the occurrence of the security breach. It is also called as sensitivity. Mathematically, it is described in equation \ref{eq:recall}:
\begin{equation}
\label{eq:recall}
\textit{Recall} = \frac{TP}{TP+FN} \times 100
\end{equation}

\textbf{F1-Score} - {} It is defined as the weighted harmonic mean of precision and recall. It tells about the ratio of correct predictions in test set. Mathematically, it is expressed in equation \ref{eq:f1score}:

\begin{equation}
\label{eq:f1score}
F1{-}Score = 2 \times \frac{Recall * Precision}{Recall + Precision}
\end{equation}

\begin{table}[!t]
\renewcommand{\arraystretch}{1.3}
\centering
\caption{Results of DDoS Attack detection on CICIDS2017\cite{sharafaldin2018toward} Dataset}
\label{results1}
\begin{tabular}{p{1.3cm} p{1.3cm} p{1.3cm} p{1.3cm} p{1.3cm}}
\hline
\textbf{Classifier}  & \textbf{Accurary} & \textbf{Precision}  & \textbf{Recall} & \textbf{F1-Score} \\  \hline
NB & 73.71 & 68.33 & 99.93 & 81.16 \\ 
KNN & 99.59 & 99.58 & 99.69 & 99.64 \\ 
RF & 100 & 100 & 100 & 100 \\ 
LR & 100 & 100 & 100 & 100 \\ \hline

\end{tabular}
\end{table}

\begin{table}[!t]
\renewcommand{\arraystretch}{1.3}
\centering
\caption{Results of Botnet detection on CTU-13 \cite{garcia2014empirical} Dataset}
\label{results2}
\begin{tabular}{p{1.3cm} p{1.3cm} p{1.3cm} p{1.3cm} p{1.3cm}}
\hline
\textbf{Classifier}  & \textbf{Accurary} & \textbf{Precision}  & \textbf{Recall} & \textbf{F1-Score} \\  \hline
NB & 71.72 & 92.89 & 36.11 & 52.01 \\ 
KNN & 97.51 & 97.67 & 96.44 & 97.05 \\ 
RF & 100 & 100 & 100 & 100 \\ 
LR & 100 & 100 & 100 & 100 \\ \hline 

\end{tabular}
\end{table}

\begin{table}[!t]
\renewcommand{\arraystretch}{1.3}
\centering
\caption{Results of Mirai Attack detection on IoT-23 \cite{iot-23} Dataset}
\label{results3}
\begin{tabular}{p{1.3cm} p{1.3cm} p{1.3cm} p{1.3cm} p{1.3cm}}
\hline
\textbf{Classifier}  & \textbf{Accurary} & \textbf{Precision}  & \textbf{Recall} & \textbf{F1-Score} \\  \hline
NB & 99.92 & 99.92 & 99.92 & 99.92 \\ 
KNN & 99.94 & 99.94 & 99.94 & 99.94 \\ 
RF & 100 & 100 & 100 & 100 \\ 
LR & 99.91 & 99.91 & 99.91 & 99.91 \\ \hline

\end{tabular}
\end{table}

Table \ref{results1} - Table \ref{results3} summarize the overall results acquired on testing the four machine learning classifiers trained over the proposed universal features set for detecting different botnet attacks across three datasets. It can be noticed that the RF classifier performed best for detecting the botnet attacks in all three datasets. On the other hand, NB classifier showed lowest performance for detecting botnet attacks in all three datasets. 

The NB classifier is based upon the Bayes' theorem with an assumption that all the features used for training the classifier are not correalted. While in our case, the features like number of forward packets per second (Fwd Pkts/s), number of backward packets per second (Bwd Pkts/s) and Down/Up Ratio are correlated with each other. Due to the existence of correlation among features, the Naive Bayes algorithm performed poor as compared to other algorithms. on the other hand, the random forest classifier consists of bagged decision trees which split on a subset of features. The random forest classifier is robust to handle the outliers, and non-linear data.

In summary, all three datasets had different types of botnet attacks, however, the proposed universal features set manifested the preeminent results for detecting the botnet attacks when trained and tested the machine learning models using the proposed universal features set across three botnet attack datasets. The experimental results reveal that the proposed universal features set efficiently support the machine algorithms to better detect the different types of botnet attacks irrespective of a dataset.


\section{Conclusion}
With the escalating adoption of insecure IoT devices, botnet attacks have become a major security threat to the internet. So far, many machine learning based solutions have been proposed for different botnet attacks detection. The performance of these machine learning based solutions primarily depends upon the features set that is used for training the machine learning models. Usually, the features selected from one certain dataset do not support machine learning models for efficiently detecting the botnet attacks on other datasets due to the diversity of the botnet attacks. Therefore, in this work, we proposed a universal features set to better support the machine learning models for detecting different botnet attacks irrespective of the dataset. The proposed features set is used to train four commonly used machine learning algorithms for detecting the botnet attacks across three different datasets. The experimental results demonstrated that the machine learning algorithms efficiently detected the botnet attacks when trained and tested these algorithms using the proposed universal features set over three different datasets. 

In the current work, we had to train the machine learning models across each dataset due to the diversity of botnet attacks. In future, we aim to generalize the machine learning models as well for detecting all types of botnet attacks.


\bibliographystyle{IEEEtran}
\bibliography{Ref.bib}

\begin{thebibliography}{10}
\providecommand{\url}[1]{#1}
\csname url@samestyle\endcsname
\providecommand{\newblock}{\relax}
\providecommand{\bibinfo}[2]{#2}
\providecommand{\BIBentrySTDinterwordspacing}{\spaceskip=0pt\relax}
\providecommand{\BIBentryALTinterwordstretchfactor}{4}
\providecommand{\BIBentryALTinterwordspacing}{\spaceskip=\fontdimen2\font plus
\BIBentryALTinterwordstretchfactor\fontdimen3\font minus
  \fontdimen4\font\relax}
\providecommand{\BIBforeignlanguage}[2]{{%
\expandafter\ifx\csname l@#1\endcsname\relax
\typeout{** WARNING: IEEEtran.bst: No hyphenation pattern has been}%
\typeout{** loaded for the language `#1'. Using the pattern for}%
\typeout{** the default language instead.}%
\else
\language=\csname l@#1\endcsname
\fi
#2}}
\providecommand{\BIBdecl}{\relax}
\BIBdecl

\bibitem{xia2020modeling}
H.~Xia, L.~Li, X.~Cheng, X.~Cheng, and T.~Qiu, ``Modeling and analysis botnet
  propagation in social internet of things,'' \emph{IEEE Internet of Things
  Journal}, 2020.

\bibitem{ghazanfar2020iot}
S.~Ghazanfar, F.~Hussain, A.~U. Rehman, U.~U. Fayyaz, F.~Shahzad, and G.~A.
  Shah, ``Iot-flock: An open-source framework for iot traffic generation,'' in
  \emph{2020 International Conference on Emerging Trends in Smart Technologies
  (ICETST)}.\hskip 1em plus 0.5em minus 0.4em\relax IEEE, 2020, pp. 1--6.

\bibitem{hossain2019application}
E.~Hossain, I.~Khan, F.~Un-Noor, S.~S. Sikander, and M.~S.~H. Sunny,
  ``Application of big data and machine learning in smart grid, and associated
  security concerns: A review,'' \emph{IEEE Access}, vol.~7, pp.
  13\,960--13\,988, 2019.

\bibitem{alladi2020consumer}
T.~Alladi, V.~Chamola, B.~Sikdar, and K.-K.~R. Choo, ``Consumer iot: Security
  vulnerability case studies and solutions,'' \emph{IEEE Consumer Electronics
  Magazine}, vol.~9, no.~2, pp. 17--25, 2020.

\bibitem{OWASP}
\BIBentryALTinterwordspacing
\emph{OWASP Releases Latest Top 10 IoT Vulnerabilities}, (accessed April 26,
  2020). [Online]. Available:
  \url{https://www.techwell.com/techwell-insights/2019/01/owasp-releases-latest-top-10-iot-vulnerabilities}
\BIBentrySTDinterwordspacing

\bibitem{ahmed2019protecting}
Z.~Ahmed, S.~M. Danish, H.~K. Qureshi, and M.~Lestas, ``Protecting iots from
  mirai botnet attacks using blockchains,'' in \emph{2019 IEEE 24th
  International Workshop on Computer Aided Modeling and Design of Communication
  Links and Networks (CAMAD)}.\hskip 1em plus 0.5em minus 0.4em\relax IEEE,
  2019, pp. 1--6.

\bibitem{pour2020data}
M.~S. Pour, A.~Mangino, K.~Friday, M.~Rathbun, E.~Bou-Harb, F.~Iqbal,
  S.~Samtani, J.~Crichigno, and N.~Ghani, ``On data-driven curation, learning,
  and analysis for inferring evolving internet-of-things (iot) botnets in the
  wild,'' \emph{Computers \& Security}, vol.~91, p. 101707, 2020.

\bibitem{bertino2017botnets}
E.~Bertino and N.~Islam, ``Botnets and internet of things security,''
  \emph{Computer}, vol.~50, no.~2, pp. 76--79, 2017.

\bibitem{gitAttack}
\BIBentryALTinterwordspacing
\emph{GitHub Survived the Biggest DDoS Attack Ever Recorded}, (accessed April
  26, 2020). [Online]. Available:
  \url{https://www.wired.com/story/github-ddos-memcached/}
\BIBentrySTDinterwordspacing

\bibitem{zarpelao2017survey}
B.~B. Zarpelao, R.~S. Miani, C.~T. Kawakani, and S.~C. de~Alvarenga, ``A survey
  of intrusion detection in internet of things,'' \emph{Journal of Network and
  Computer Applications}, vol.~84, pp. 25--37, 2017.

\bibitem{barnes2017assessing}
J.~Barnes, R.~Klinger, and S.~S. im~Walde, ``Assessing state-of-the-art
  sentiment models on state-of-the-art sentiment datasets,'' in
  \emph{Proceedings of the 8th Workshop on Computational Approaches to
  Subjectivity, Sentiment and Social Media Analysis}, 2017, pp. 2--12.

\bibitem{frank2018protecting}
C.~Frank, C.~Nance, S.~Jarocki, and W.~E. Pauli, ``Protecting iot from mirai
  botnets; iot device hardening,'' \emph{Journal of Information Systems Applied
  Research}, vol.~11, no.~2, p.~33, 2018.

\bibitem{ryu2018comparative}
S.~Ryu, B.~Yang \emph{et~al.}, ``A comparative study of machine learning
  algorithms and their ensembles for botnet detection,'' \emph{Journal of
  Computer and Communications}, vol.~6, no.~05, p. 119, 2018.

\bibitem{kato2017development}
K.~Kato and V.~Klyuev, ``Development of a network intrusion detection system
  using apache hadoop and spark,'' in \emph{2017 IEEE Conference on Dependable
  and Secure Computing}.\hskip 1em plus 0.5em minus 0.4em\relax IEEE, 2017, pp.
  416--423.

\bibitem{lagraa2017botgm}
S.~Lagraa, J.~Fran{\c{c}}ois, A.~Lahmadi, M.~Miner, C.~Hammerschmidt, and
  R.~State, ``Botgm: Unsupervised graph mining to detect botnets in traffic
  flows,'' in \emph{2017 1st Cyber Security in Networking Conference
  (CSNet)}.\hskip 1em plus 0.5em minus 0.4em\relax IEEE, 2017, pp. 1--8.

\bibitem{wang2017botnet}
J.~Wang and I.~C. Paschalidis, ``Botnet detection based on anomaly and
  community detection,'' \emph{IEEE Transactions on Control of Network
  Systems}, vol.~4, no.~2, 2017.

\bibitem{daya2019graph}
A.~A. Daya, M.~A. Salahuddin, N.~Limam, and R.~Boutaba, ``A graph-based machine
  learning approach for bot detection,'' in \emph{2019 IFIP/IEEE Symposium on
  Integrated Network and Service Management (IM)}.\hskip 1em plus 0.5em minus
  0.4em\relax IEEE, 2019, pp. 144--152.

\bibitem{piskozub2019malalert}
M.~Piskozub, R.~Spolaor, and I.~Martinovic, ``Malalert: Detecting malware in
  large-scale network traffic using statistical features,'' \emph{ACM
  SIGMETRICS Performance Evaluation Review}, vol.~46, no.~3, pp. 151--154,
  2019.

\bibitem{chen2020novel}
T.~Chen, G.~Zhou, Z.~Liu, and T.~Jing, ``A novel ensemble anomaly based
  approach for command and control channel detection,'' in \emph{Proceedings of
  the 2020 4th International Conference on Cryptography, Security and Privacy},
  2020, pp. 74--78.

\bibitem{blaise2020botfp}
A.~Blaise, M.~Bouet, V.~Conan, and S.~Secci, ``Botfp: Fingerprints clustering
  for bot detection,'' in \emph{IEEE/IFIP Network Operations and Management
  Symposium (NOMS)}, 2020.

\bibitem{letteri2019feature}
I.~Letteri, G.~Della~Penna, and P.~Caianiello, ``Feature selection strategies
  for http botnet traffic detection,'' in \emph{2019 IEEE European Symposium on
  Security and Privacy Workshops (EuroS\&PW)}.\hskip 1em plus 0.5em minus
  0.4em\relax IEEE, 2019, pp. 202--210.

\bibitem{guerra2019hybrid}
A.~Guerra-Manzanares, H.~Bahsi, and S.~N{\~o}mm, ``Hybrid feature selection
  models for machine learning based botnet detection in iot networks,'' in
  \emph{2019 International Conference on Cyberworlds (CW)}.\hskip 1em plus
  0.5em minus 0.4em\relax IEEE, 2019, pp. 324--327.

\bibitem{sharafaldin2018toward}
I.~Sharafaldin, A.~H. Lashkari, and A.~A. Ghorbani, ``Toward generating a new
  intrusion detection dataset and intrusion traffic characterization.'' in
  \emph{4th International Conference on Information Systems Security and
  Privacy (ICISSP)}, 2018, pp. 108--116.

\bibitem{garcia2014empirical}
S.~Garcia, M.~Grill, J.~Stiborek, and A.~Zunino, ``An empirical comparison of
  botnet detection methods,'' \emph{computers \& security}, vol.~45, pp.
  100--123, 2014.

\bibitem{iot-23}
\BIBentryALTinterwordspacing
\emph{Stratosphere Laboratory. A labeled dataset with malicious and benign IoT
  network traffic. Agustin Parmisano, Sebastian Garcia, Maria Jose Erquiaga},
  (accessed April 26, 2020). [Online]. Available:
  \url{https://www.stratosphereips.org/datasets-iot23}
\BIBentrySTDinterwordspacing

\bibitem{CICFLOWMETER}
\BIBentryALTinterwordspacing
\emph{CICFLOWMETER}, (accessed June 30, 2020). [Online]. Available:
  \url{https://github.com/ahlashkari/CICFlowMeter}
\BIBentrySTDinterwordspacing

\bibitem{tcpdump}
\BIBentryALTinterwordspacing
\emph{WinDump: tcpdump for Windows using WinPcap}, (accessed May 18, 2020).
  [Online]. Available: \url{https://www.winpcap.org/windump/}
\BIBentrySTDinterwordspacing

\bibitem{cicflowmeter_features}
\BIBentryALTinterwordspacing
\emph{NETWORK TRAFFIC FLOW ANALYZER}, (accessed January 6, 2020). [Online].
  Available: \url{http://www.netflowmeter.ca/netflowmeter.html}
\BIBentrySTDinterwordspacing

\bibitem{stamos2015learning}
D.~Stamos, S.~Martelli, M.~Nabi, A.~McDonald, V.~Murino, and M.~Pontil,
  ``Learning with dataset bias in latent subcategory models,'' in
  \emph{Proceedings of the IEEE Conference on Computer Vision and Pattern
  Recognition}, 2015, pp. 3650--3658.

\bibitem{dedeturk2020spam}
B.~K. Dedeturk and B.~Akay, ``Spam filtering using a logistic regression model
  trained by an artificial bee colony algorithm,'' \emph{Applied Soft
  Computing}, p. 106229, 2020.

\end{thebibliography}

\end{document}